\documentclass[aip, apl, amsmath,amssymb,reprint,graphicx,10pt]{revtex4-1}
	\usepackage{bm}
	\usepackage{siunitx}
	\usepackage{pstool} 
	\usepackage{epstopdf}
	
\draft
\begin{document}

\title{A linear triple quantum dot system in isolated configuration}

\author{Hanno Flentje}
\author{Benoit Bertrand}
\author{Pierre-Andr\'e Mortemousque}
\author{Vivien Thiney}
\affiliation{Institut N\'eel, CNRS and Universit\'e Grenoble Alpes, 38042 Grenoble, France}
\author{Arne Ludwig}
\author{Andreas D. Wieck}
\affiliation{Lehrstuhl f\"ur Angewandte Festk\"orperphysik, Ruhr-Universit\"at Bochum, Universit\"atsstra{\ss}e 150, 44780 Bochum, Germany}
\author{Christopher B\"auerle}
\affiliation{Institut N\'eel, CNRS and Universit\'e Grenoble Alpes, 38042 Grenoble, France}
\author{Tristan Meunier}
\affiliation{Institut N\'eel, CNRS and Universit\'e Grenoble Alpes, 38042 Grenoble, France}

\date{\today}
\begin{abstract}
The scaling up of electron spin qubit based nanocircuits has remained challenging up to date and involves the development of efficient charge control strategies. Here we report on the experimental realization of a linear triple quantum dot in a regime isolated from the reservoir. We show how this regime can be reached with a fixed number of electrons. Charge stability diagrams of the one, two and three electron configurations where only electron exchange between the dots is allowed are observed. They are modelled with established theory based on a capacitive model of the dot systems. The advantages of the isolated regime with respect to experimental realizations of quantum simulators and qubits are discussed. We envision that the results presented here will make more manipulation schemes for existing qubit implementations possible and will ultimately allow to increase the number of tunnel coupled quantum dots which can be simultaneously controlled.
\end{abstract}

\keywords{multi quantum dot systems, semiconductor nanostructures, single charge detection}

\maketitle

Electrons trapped in laterally defined quantum dots (QD) have emerged as a versatile platform for qubit implementations\cite{Loss1998, Hanson2007, Shulman2012, zwanenburg2013} as well as mesoscopic physics test-beds,\cite{goldhaber1998, Cronenwett1998,Jeong2001} and have recently attracted attention as a possible candidate for quantum simulators.\cite{Barthelemy2013} Single electron charges and their spin can nowadays be routinely trapped, manipulated and read out in single- and double-QD systems.\cite{Hanson2007, Veldhorst2014} An important experimental effort has been carried out on multidot systems and charge control of arrays made of up to four tunnel-QDs has been demonstrated. \cite{Studenikin2006,Schroeer2007,Thalineau2012,Takakura2014,Eng2015,Baart2016} As far as spin is concerned, a triangular triple QD showed charge frustration in transport measurements,\cite{Seo2013} hinting at the possibility of studying spin frustration in such a system and coherent oscillations of three spin states were demonstrated in a linear triple QD.\cite{Laird2010,Gaudreau2012,noiri2016} Despite the inherent scaling properties of lithographically defined semiconductor based systems, the coupling of the QDs with the electron reservoir implies an infinite number of possible charge configurations. The resulting complexity in the dot array tunability limits the capabilities to control simultaneously an increasing number of tunnel-coupled quantum dots and represents an important challenge for large-scale spin-based quantum information processing. 

In this study, we apply our recently developed technique of isolated charge manipulation\cite{Bertrand2015,flentje2017coherent} to a linear chain of three quantum dots. In this regime, the coupling to the electron reservoir can be neglected and only interdot charge transitions are allowed. As a result of the reduced complexity, all possible charge configurations at a fixed electron number are easily accessible and the tunnel-coupling between the dots can be controlled over several orders of magnitude \textit{in situ} while keeping the electron number fixed. We show how we can bring the linear-triple dot system into an isolated regime, and that the electron number can be prepared deterministically and kept for an arbitrarily long time. The observed charge stability diagrams are analyzed with a model based on the established constant interaction model \cite{RevModPhys.75.1} and one obtains a good agreement between theory and experiment. These results provide a guideline for scaling up the number of tunnel-coupled QDs, opening up manipulation schemes for qubit implementations, and presenting pathways for the realization of quantum simulators with quantum dot arrays.

\begin{figure}[htbp]
	\centering
	\includegraphics{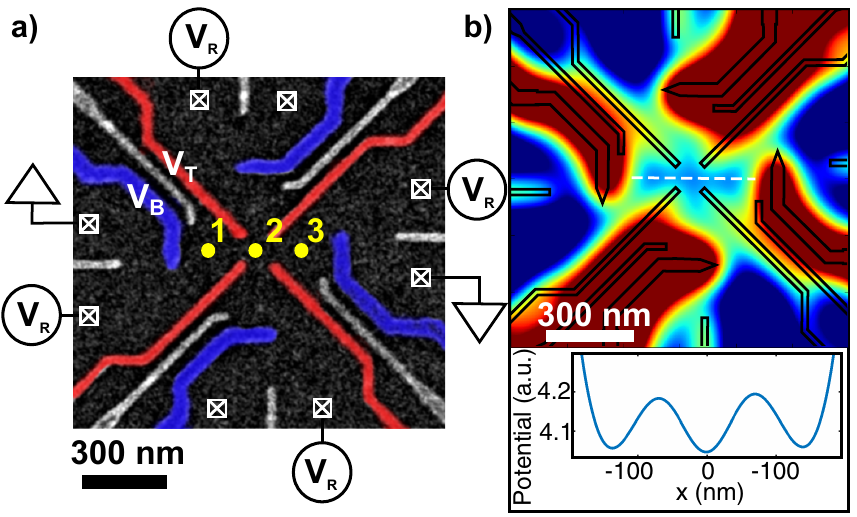}
	\caption{(a) Scanning electron microscope (SEM) image of the linear-triple dot sample. Depletion gates have been coloured for clarity. The yellow dots indicate the approximate positions of the QDs. (b) Top: Electrostatic simulation of the potential distribution with realistic gate voltages. The potential is color coded from low potential (blue) to high potential (red). The voltages applied on the top and bottom barrier gates (blue in SEM image) are more negative than on the left/right gates to achieve a linear chain of QDs. Screening effects and the random impurity distribution are not taken into account. Bottom: 2-dimensional cut along the white dashed line through the simulated potential showing the three distinct potential minima.}
	\label{fig:4dot}
\end{figure}

Our device was fabricated using a Si doped AlGaAs/GaAs heterostructure grown by molecular beam epitaxy, with a two-dimensional electron gas (2DEG) 100 \si{\nm} below the crystal surface which has a carrier mobility of $10^6$ \si{\square\cm\per\V\per\second} and an electron density of $2.7\times 10^{11}$ \si{\per\square\cm}. QDs are defined and controlled by the application of negative voltages on Ti/Au Schottky gates deposited on the surface of the crystal. A schematic view of the linear-triple dot sample can be seen in Fig. \ref{fig:4dot}(a). By applying very negative voltages on diagonally opposite bottom and top blue gates, we can tune the sample into a linear chain of three QDs as indicated in the electrostatic potential calculation of Fig.~\ref{fig:4dot}(b).\cite{Davies1995,Bautze2014} Home-made electronics ensure fast changes of both chemical potentials and tunnel-couplings with voltage pulse rise times below 2~\si{\us}. The charge configuration can be read out by four quantum point contacts (QPC), tuned to be sensitive local electrometers.\cite{Field1993} The sample is mounted in a home-made dilution cryostat with an electron temperature of 150~m\si{\K}, estimated from Coulomb peak broadening.\cite{Foxman1994,Meir1991} 

\begin{figure}[htb]
	\includegraphics[width=0.95\columnwidth]{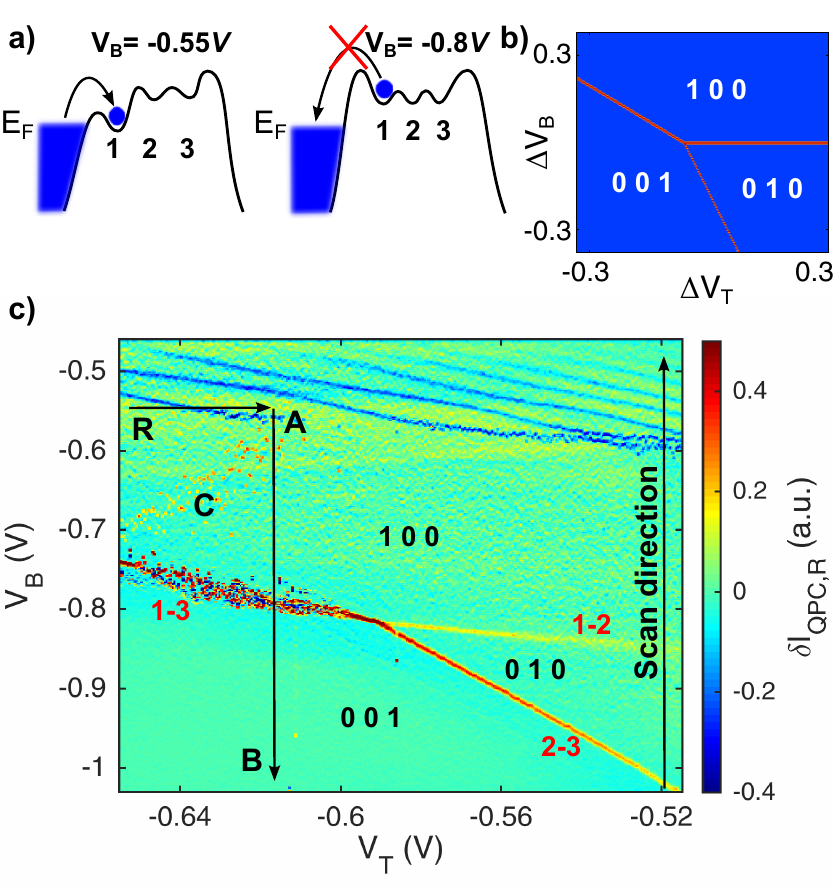}
	\caption{(a) Schematic of the QD potential with respect to the chemical potential of the 2DEG ($\mathrm{E_F}$) in the loading position (left) and isolated state (right). (b) Calculated stability diagram from the simulation for one electron loaded in the dot system. Only boundaries between charge configurations are represented as a function of $\Delta V_T$ and $\Delta V_L$. The absolute voltages are defined by the charge offset $N_0$. $V_B$ is assumed to only weakly couple to the chemical potential of QD 2 and mostly affects QD 1 while $V_T$ is assumed to couple to both chemical potentials equally. The chemical potential of QD 3 is assumed to be independent of the varied gates. (c) Differential conductance of the right QPC current with respect to $V_B$ (stability diagram). Once in the isolated configuration, $V_B$ is scanned from more negative to more positive voltages (indicated by arrow). The reset-load-isolation sequence is performed before each gate scan to set the initial electron number. Lines indicate abrupt changes in QPC conductance, where blue (red) color indicates electrons displaced closer to (farther from) the sensing QPC. The position of the electron among the three QDs is indicated in the graph following the notation in the text. In red we label the different transitions indicating the connected QDs. When $V_B$ is slowly pushed towards more positive values, the QD trapping potential becomes too shallow with respect to the barrier height, resulting in a tunneling event of the electron to the reservoirs. This manifests in stochastically appearing electron loss events (red dotted lines in the stability diagram close to the region C) which  delimit the isolated configuration with the electron number fixed to a single electron.}
	\label{fig:oneelectron}
\end{figure}

The isolation of the electrons is achieved by increasing the tunnel barriers to the leads with more negative voltages on the barrier gates (blue in Fig. \ref{fig:4dot}(a)). To access the isolated regime, a specific voltage sequence needs to be performed. We first reset the number of electrons in the QD 1 to zero at the position R (see Fig. \ref{fig:oneelectron}(c)), and then load the QD 1 in a position of high tunneling with the leads, close to the point A. By changing the position A, we can load the quantum dot with any number of electron. The system is then disconnected from the leads with a fast \si{\micro\second} voltage pulse on $V_B$ towards point B (black arrows in Fig.~\ref{fig:oneelectron}(c)) which rapidly raises the potential barrier between the quantum dot and the electron reservoir such that the electron number is preserved. A schematic of the potential at the points A and B is shown in Fig.~\ref{fig:oneelectron}(a). Starting from point B, the parameter space is explored by varying the chemical potentials of the individual QDs with $V_B$ and $V_T$. More precisely, for different $V_T$, $V_B$ is scanned from negative to positive gate voltages. For $V_B$ more positive than -0.6V, electron exchange between the dots and the reservoir is possible and leads to changes of electron number in the dot, which are detected as peaks in the differential conductance of the electrometer. This corresponds to the blue charge degeneracy lines in the stability diagram as can be seen in the top of Fig.~\ref{fig:oneelectron}(c). These lines separate the charge regions of QD1 and are used to calibrate where to tune the point A to load an arbitrary number of electrons. For $V_B$ more negative than -0.6V, the charge degeneracy line disappears, showing that the dwell time of an electron in the QD becomes much larger than the measurement time and the system becomes effectively decoupled from the reservoir. In other terms, the only allowed charge transitions for electrons are between adjacent QDs. These interdot electron tunnel events will change the detector QPC current similarly to classical charge degeneracy lines and can therefore be detected as shown in the lower part of Fig.~\ref{fig:oneelectron}(c). We note that the timescale for the $V_B$-scan is several seconds, confirming that an electron can be kept isolated in the QD for a long duration. The result of the isolation procedure manifests itself directly on the stability diagram presented in Fig. \ref{fig:oneelectron}(c).

In the case of one-electron-loading, we observe three distinct degeneracy lines indicating charge transitions between QDs. Their slopes permit to label the three obtained charge configurations. For more positive voltages on the gates $V_B$ and $V_T$ (see Fig. \ref{fig:4dot}), the electron is confined in the loading QD 1 (see Fig. \ref{fig:oneelectron}(a)). We label this configuration (1 0 0) indicating the number of electrons in the respective QDs. By making the voltage on $V_B$ more negative, it is possible to move the electron into QD 2 (0 1 0). For very negative voltages on both gates, the electron is moved to QD 3 (0 0 1). As expected from the sample geometry, the voltage $V_T$ strongly affects the height of the energy barrier between QD 1 and QD 2 and therefore the tunnel-coupling between the respective QDs. For positive values of $V_T$, this increased tunnel-coupling leads to a broadening of the associated degeneracy line.\cite{DiCarlo2004} Conversely, the QD 2- QD 3 degeneracy line is almost unaffected by $V_T$. Due to linear configuration of the dot array, the electron tunneling process between QD 1 and QD 3 is the result of an indirect coupling mediated by the QD 2.\cite{braakman2013long,Takakura2014} As a consequence, we expect a strong dependence of the associated energy with the two direct tunnel-couplings (QD 1-QD 2 and QD 2-QD 3) as well as the energy detuning with the chemical potential of QD 2. In the geometry of the sample, both the QD 1-QD 2 coupling and the detuning are controlled by $V_T$. It explains the observed fast change of the QD 1-QD 3 degeneracy line shape suggesting the QD 1-QD 3 tunnel-coupling going to zero for $V_T$ smaller than -0.6V.

Despite these strong changes in the tunneling rate, the transition lines in the stability diagram stay clearly visible over a large parameter range showing the insensitivity of this approach to an initially unknown coupling strength. It nevertheless disappears in two regimes: when the tunneling time becomes much longer than the measurement time such that the events become stochastic ($\approx$Hz; left of $V_T\approx-0.6 V, V_B\approx-0.8 V$ in Fig.~\ref{fig:oneelectron}(c)). Secondly, it disappears in the limit where the tunneling energy becomes much larger than temperature ($\approx$GHz), and therefore the line broadens until no longer being observable (right part of Fig.~\ref{fig:oneelectron}(c)). Moreover, the QD 1-QD 2 tunnel-coupling can then be continuously varied by changing $V_T$ without erroneously changing the electron number. This controlled variability of the tunnel-coupling in the isolated regime has recently been used to implement a spin manipulation scheme which is partially protected from charge noise.\cite{Bertrand2015}



\begin{figure}[tb]
	\includegraphics{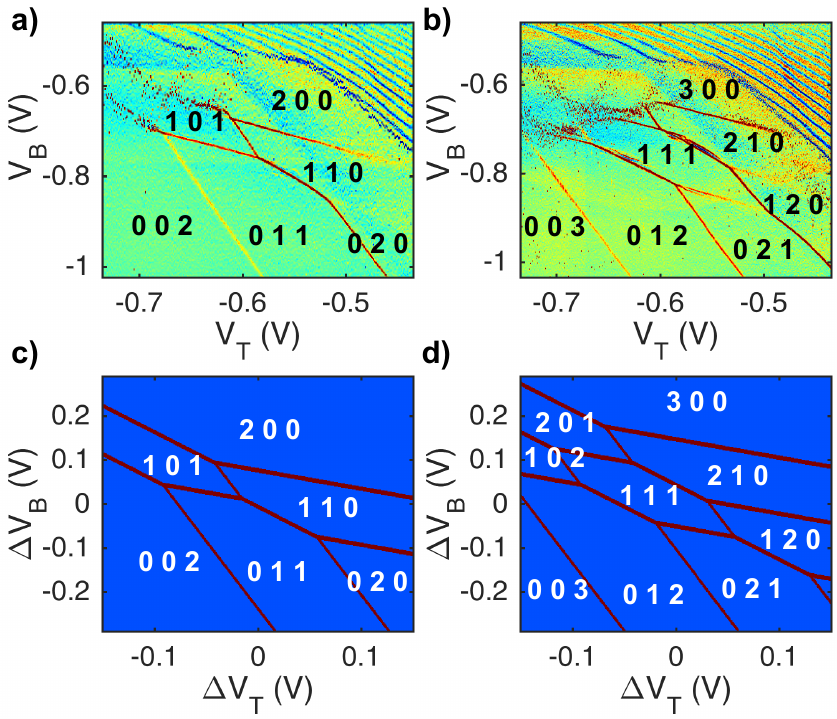}
	\caption{Stability diagrams in the isolated regime for two (a) and three (b) electrons respectively with a procedure identical to Fig.~\ref{fig:oneelectron} and with the same color coding of the derivative of the detector current. A controlled number of electrons is first loaded and then $V_B$ is scanned from negative to positive voltages. Calculated stability diagrams from the simulation for two (c) and three (d) electrons loaded in the dot system. Only boundaries between charge configurations are represented as a function of $\Delta V_T$ and $\Delta V_L$ (see caption of Fig.~\ref{fig:oneelectron}b).}
	\label{fig:ChargeDiagrams}
\end{figure}

To show the straightforwardness of this approach for scaling-up the occupation number, we load the triple QD with two or three electrons, and perform the same gate scans (Fig.~\ref{fig:ChargeDiagrams}(a,b)). Similar to the one electron case, the isolated degeneracy lines are characterized by the same three slopes and the different charge configurations are therefore assigned as represented in Fig \ref{fig:oneelectron}. 

\begin{figure}[b]
	\vspace*{2ex}
	\includegraphics{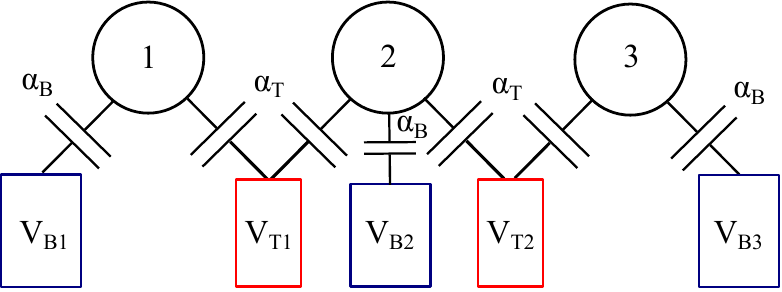}
	\caption{Schematic of the electrostatic model used to simulate the response of the system to variations of the gate voltages. The change of chemical potentials of different QDs are linked to the gate voltages by coefficients $\alpha_x$. For the simulations we used $\vec{\alpha}_{B}$=(0.05 0.0202 0)$e$, $\vec{\alpha}_{T}$=(0.0709 0.0563 0)$e$ and $\vec{E}_C$=(2.4 2.1 2.6) \si{\milli\electronvolt}}
	\label{fig:theory_scheme}
\end{figure}

Furthermore, we can quantitatively understand the structure formed by the transition lines with the constant interaction model.\cite{VanHouten1992,Hanson2007} In the electrostatic model presented in Fig.~\ref{fig:theory_scheme}, the coupling between electrons in the QDs and the gates is modelled as a sum of capacitances.  We first introduce the renormalized gate capacitances $\alpha_{jx}$ to parametrize the effect of gate voltage $V_j$ on the potential of QD $x$. Then the energy of the charge configurations $\vec{k}_{i}$ (i.e. for $N=1$, $\vec{k}_{1}=$(1 0 0)) is given by \cite{VanHouten1992}

\begin{equation}
 \label{eq:test}
E_{\vec{k}_{i}}= \sum_x \frac{1}{E_{C,x}}\{\sum_j \alpha_{jx} V_j+[k_{ix}-N_0] E_{C,x}\}^2,
\end{equation}

\noindent where the summation is over all QDs $x$, $E_{C,x}$ is the charging energy of QD $x$ and $N_0$ is the number of electrons compensating the positive donor charges which are related to the electron density of the 2DEG. As the electron reservoirs can no longer be used as chemical potential references, ${k_{i}}$ has to be limited to the subspace of configurations with constant total electron charges and the smallest $E_{k_{i}}$ will be the ground state. In the model, we neglect the finite capacitive coupling between QDs and the dot orbital energies. These effects do not change the shape of the diagram, but only renormalize the involved capacitances and charging energies. The boundaries between the energetically lowest lying charge configurations as a function of the gate voltages has been plotted for different total number of electrons in Fig.~\ref{fig:oneelectron}(b) and Fig.~\ref{fig:ChargeDiagrams}(c,d) and show an excellent qualitative agreement with the experimental data. At a degeneracy line the energy of the two associated quantum dot configurations is equal. The respective slopes of the transitions in the experimental diagrams therefore allow to infer the ratios $\alpha_T$/$\alpha_B$ for the respective QDs. The absolute values can in principle be measured using transport measurements or photon-assisted tunneling\cite{Oosterkamp1998} (possible in the isolated regime). Here we got quantitative agreement by guessing one parameter ($\alpha_{B1}$) and neglecting any influence of $V_B$ and $V_T$ on QD 3 (see Fig.~\ref{fig:theory_scheme}). The charging energies of the QDs are determined by the positions of the parallel charge degeneracy line crossings in the diagram with two electrons (Fig.~\ref{fig:ChargeDiagrams}(a)). 



The reduced complexity of the obtained charge diagrams can be harnessed for the operation of large quantum dot systems. As electrons can access all configurations without unwanted electron exchange with the reservoirs due to pulse imperfections, this approach of fixed electron number manipulation allows to increase the effective size of the available configuration space. The isolation also allows to study the behavior of systems in which an increasing number of QDs prevents the more distant leads to be used as an effective chemical potential reference.

In conclusion, we have performed a full control sequence for a multidot system decoupled from the leads. We have demonstrated how the system can be initialized with a desired number of electrons, and that all possible charge configurations can be easily accessed with the same electrons. The resulting system can be understood and modelled with established theory and shows qualitative and quantitative consistency with the measured diagrams. The reduced complexity of the isolated regime makes charge reorganization in the structure straightforward and permit higher tunability of the dot parameters. We conclude that this approach will allow to increase the size of multiqubit systems as well as open up manipulation schemes for existing systems.  Finally, the concept of isolated charge manipulation can be directly extended to similar qubit architectures such as electron spins in silicon, and should therefore find wide application in future experiments.


%
%
%
%
%


\begin{acknowledgments}
We acknowledge technical support from the ``Poles
Electroniques" of the ``D\'{e}partement Nano and MCBT"
from the Institut N\'{e}el as well as Henry Rodenas for technical support. A.D.W. acknowledges support of the DFG SPP1285 and
the BMBF QuaHLRep 01BQ1035. T.M. acknowledges
financial support ERC "QSPINMOTION" (307149). We are
grateful to the Nanoscience Foundation of Grenoble, for
partial financial support of this work. Devices were fabricated at ``Plateforme Technologique Amont" de Grenoble, with the financial support of the ``Nanosciences aux
limites de la Nano\'{e}lectronique" Foundation and CNRS
Renatech network. 
\end{acknowledgments}

%





\begin{thebibliography}{31}%
\makeatletter
\providecommand \@ifxundefined [1]{%
 \@ifx{#1\undefined}
}%
\providecommand \@ifnum [1]{%
 \ifnum #1\expandafter \@firstoftwo
 \else \expandafter \@secondoftwo
 \fi
}%
\providecommand \@ifx [1]{%
 \ifx #1\expandafter \@firstoftwo
 \else \expandafter \@secondoftwo
 \fi
}%
\providecommand \natexlab [1]{#1}%
\providecommand \enquote  [1]{``#1''}%
\providecommand \bibnamefont  [1]{#1}%
\providecommand \bibfnamefont [1]{#1}%
\providecommand \citenamefont [1]{#1}%
\providecommand \href@noop [0]{\@secondoftwo}%
\providecommand \href [0]{\begingroup \@sanitize@url \@href}%
\providecommand \@href[1]{\@@startlink{#1}\@@href}%
\providecommand \@@href[1]{\endgroup#1\@@endlink}%
\providecommand \@sanitize@url [0]{\catcode `\\12\catcode `\$12\catcode
  `\&12\catcode `\#12\catcode `\^12\catcode `\_12\catcode `\%12\relax}%
\providecommand \@@startlink[1]{}%
\providecommand \@@endlink[0]{}%
\providecommand \url  [0]{\begingroup\@sanitize@url \@url }%
\providecommand \@url [1]{\endgroup\@href {#1}{\urlprefix }}%
\providecommand \urlprefix  [0]{URL }%
\providecommand \Eprint [0]{\href }%
\providecommand \doibase [0]{http://dx.doi.org/}%
\providecommand \selectlanguage [0]{\@gobble}%
\providecommand \bibinfo  [0]{\@secondoftwo}%
\providecommand \bibfield  [0]{\@secondoftwo}%
\providecommand \translation [1]{[#1]}%
\providecommand \BibitemOpen [0]{}%
\providecommand \bibitemStop [0]{}%
\providecommand \bibitemNoStop [0]{.\EOS\space}%
\providecommand \EOS [0]{\spacefactor3000\relax}%
\providecommand \BibitemShut  [1]{\csname bibitem#1\endcsname}%
\let\auto@bib@innerbib\@empty
\bibitem [{\citenamefont {Loss}\ and\ \citenamefont
  {DiVincenzo}(1998)}]{Loss1998}%
  \BibitemOpen
  \bibfield  {author} {\bibinfo {author} {\bibfnamefont {D.}~\bibnamefont
  {Loss}}\ and\ \bibinfo {author} {\bibfnamefont {D.~P.}\ \bibnamefont
  {DiVincenzo}},\ }\href {\doibase 10.1103/PhysRevA.57.120} {\bibfield
  {journal} {\bibinfo  {journal} {Phys. Rev. A}\ }\textbf {\bibinfo {volume}
  {57}},\ \bibinfo {pages} {120} (\bibinfo {year} {1998})}\BibitemShut
  {NoStop}%
\bibitem [{\citenamefont {Hanson}\ \emph {et~al.}(2007)\citenamefont {Hanson},
  \citenamefont {Kouwenhoven}, \citenamefont {Petta}, \citenamefont {Tarucha},\
  and\ \citenamefont {Vandersypen}}]{Hanson2007}%
  \BibitemOpen
  \bibfield  {author} {\bibinfo {author} {\bibfnamefont {R.}~\bibnamefont
  {Hanson}}, \bibinfo {author} {\bibfnamefont {L.~P.}\ \bibnamefont
  {Kouwenhoven}}, \bibinfo {author} {\bibfnamefont {J.~R.}\ \bibnamefont
  {Petta}}, \bibinfo {author} {\bibfnamefont {S.}~\bibnamefont {Tarucha}}, \
  and\ \bibinfo {author} {\bibfnamefont {L.~M.~K.}\ \bibnamefont
  {Vandersypen}},\ }\href {\doibase 10.1103/RevModPhys.79.1217} {\bibfield
  {journal} {\bibinfo  {journal} {Rev. Mod. Phys.}\ }\textbf {\bibinfo {volume}
  {79}},\ \bibinfo {pages} {1217} (\bibinfo {year} {2007})}\BibitemShut
  {NoStop}%
\bibitem [{\citenamefont {Shulman}\ \emph {et~al.}(2012)\citenamefont
  {Shulman}, \citenamefont {Dial}, \citenamefont {Harvey}, \citenamefont
  {Bluhm}, \citenamefont {Umansky},\ and\ \citenamefont
  {Yacoby}}]{Shulman2012}%
  \BibitemOpen
  \bibfield  {author} {\bibinfo {author} {\bibfnamefont {M.~D.}\ \bibnamefont
  {Shulman}}, \bibinfo {author} {\bibfnamefont {O.~E.}\ \bibnamefont {Dial}},
  \bibinfo {author} {\bibfnamefont {S.~P.}\ \bibnamefont {Harvey}}, \bibinfo
  {author} {\bibfnamefont {H.}~\bibnamefont {Bluhm}}, \bibinfo {author}
  {\bibfnamefont {V.}~\bibnamefont {Umansky}}, \ and\ \bibinfo {author}
  {\bibfnamefont {A.}~\bibnamefont {Yacoby}},\ }\href {\doibase
  10.1126/science.1217692} {\bibfield  {journal} {\bibinfo  {journal}
  {Science}\ }\textbf {\bibinfo {volume} {336}},\ \bibinfo {pages} {202}
  (\bibinfo {year} {2012})}\BibitemShut {NoStop}%
\bibitem [{\citenamefont {Zwanenburg}\ \emph {et~al.}(2013)\citenamefont
  {Zwanenburg}, \citenamefont {Dzurak}, \citenamefont {Morello}, \citenamefont
  {Simmons}, \citenamefont {Hollenberg}, \citenamefont {Klimeck}, \citenamefont
  {Rogge}, \citenamefont {Coppersmith},\ and\ \citenamefont
  {Eriksson}}]{zwanenburg2013}%
  \BibitemOpen
  \bibfield  {author} {\bibinfo {author} {\bibfnamefont {F.~A.}\ \bibnamefont
  {Zwanenburg}}, \bibinfo {author} {\bibfnamefont {A.~S.}\ \bibnamefont
  {Dzurak}}, \bibinfo {author} {\bibfnamefont {A.}~\bibnamefont {Morello}},
  \bibinfo {author} {\bibfnamefont {M.~Y.}\ \bibnamefont {Simmons}}, \bibinfo
  {author} {\bibfnamefont {L.~C.}\ \bibnamefont {Hollenberg}}, \bibinfo
  {author} {\bibfnamefont {G.}~\bibnamefont {Klimeck}}, \bibinfo {author}
  {\bibfnamefont {S.}~\bibnamefont {Rogge}}, \bibinfo {author} {\bibfnamefont
  {S.~N.}\ \bibnamefont {Coppersmith}}, \ and\ \bibinfo {author} {\bibfnamefont
  {M.~A.}\ \bibnamefont {Eriksson}},\ }\href@noop {} {\bibfield  {journal}
  {\bibinfo  {journal} {Reviews of modern physics}\ }\textbf {\bibinfo {volume}
  {85}},\ \bibinfo {pages} {961} (\bibinfo {year} {2013})}\BibitemShut
  {NoStop}%
\bibitem [{\citenamefont {Goldhaber-Gordon}\ \emph {et~al.}(1998)\citenamefont
  {Goldhaber-Gordon}, \citenamefont {Shtrikman}, \citenamefont {Mahalu},
  \citenamefont {Abusch-Magder}, \citenamefont {Meirav},\ and\ \citenamefont
  {Kastner}}]{goldhaber1998}%
  \BibitemOpen
  \bibfield  {author} {\bibinfo {author} {\bibfnamefont {D.}~\bibnamefont
  {Goldhaber-Gordon}}, \bibinfo {author} {\bibfnamefont {H.}~\bibnamefont
  {Shtrikman}}, \bibinfo {author} {\bibfnamefont {D.}~\bibnamefont {Mahalu}},
  \bibinfo {author} {\bibfnamefont {D.}~\bibnamefont {Abusch-Magder}}, \bibinfo
  {author} {\bibfnamefont {U.}~\bibnamefont {Meirav}}, \ and\ \bibinfo {author}
  {\bibfnamefont {M.}~\bibnamefont {Kastner}},\ }\href@noop {} {\bibfield
  {journal} {\bibinfo  {journal} {Nature}\ }\textbf {\bibinfo {volume} {391}},\
  \bibinfo {pages} {156} (\bibinfo {year} {1998})}\BibitemShut {NoStop}%
\bibitem [{\citenamefont {Cronenwett}, \citenamefont {Oosterkamp},\ and\
  \citenamefont {Kouwenhoven}(1998)}]{Cronenwett1998}%
  \BibitemOpen
  \bibfield  {author} {\bibinfo {author} {\bibfnamefont {S.~M.}\ \bibnamefont
  {Cronenwett}}, \bibinfo {author} {\bibfnamefont {T.~H.}\ \bibnamefont
  {Oosterkamp}}, \ and\ \bibinfo {author} {\bibfnamefont {L.~P.}\ \bibnamefont
  {Kouwenhoven}},\ }\href {\doibase 10.1126/science.281.5376.540} {\bibfield
  {journal} {\bibinfo  {journal} {Science}\ }\textbf {\bibinfo {volume}
  {281}},\ \bibinfo {pages} {540} (\bibinfo {year} {1998})}\BibitemShut
  {NoStop}%
\bibitem [{\citenamefont {Jeong}, \citenamefont {Chang},\ and\ \citenamefont
  {Melloch}(2001)}]{Jeong2001}%
  \BibitemOpen
  \bibfield  {author} {\bibinfo {author} {\bibfnamefont {H.}~\bibnamefont
  {Jeong}}, \bibinfo {author} {\bibfnamefont {A.~M.}\ \bibnamefont {Chang}}, \
  and\ \bibinfo {author} {\bibfnamefont {M.~R.}\ \bibnamefont {Melloch}},\
  }\href {\doibase 10.1126/science.1063182} {\bibfield  {journal} {\bibinfo
  {journal} {Science}\ }\textbf {\bibinfo {volume} {293}},\ \bibinfo {pages}
  {2221} (\bibinfo {year} {2001})}\BibitemShut {NoStop}%
\bibitem [{\citenamefont {Barthelemy}\ and\ \citenamefont
  {Vandersypen}(2013)}]{Barthelemy2013}%
  \BibitemOpen
  \bibfield  {author} {\bibinfo {author} {\bibfnamefont {P.}~\bibnamefont
  {Barthelemy}}\ and\ \bibinfo {author} {\bibfnamefont {L.~M.~K.}\ \bibnamefont
  {Vandersypen}},\ }\href {\doibase 10.1002/andp.201300124} {\bibfield
  {journal} {\bibinfo  {journal} {Ann. Phys.}\ }\textbf {\bibinfo {volume}
  {525}},\ \bibinfo {pages} {808} (\bibinfo {year} {2013})}\BibitemShut
  {NoStop}%
\bibitem [{\citenamefont {Veldhorst}\ \emph {et~al.}(2014)\citenamefont
  {Veldhorst}, \citenamefont {Hwang}, \citenamefont {Yang}, \citenamefont
  {Leenstra}, \citenamefont {De~Ronde}, \citenamefont {Dehollain},
  \citenamefont {Muhonen}, \citenamefont {Hudson}, \citenamefont {Itoh},
  \citenamefont {Morello} \emph {et~al.}}]{Veldhorst2014}%
  \BibitemOpen
  \bibfield  {author} {\bibinfo {author} {\bibfnamefont {M.}~\bibnamefont
  {Veldhorst}}, \bibinfo {author} {\bibfnamefont {J.}~\bibnamefont {Hwang}},
  \bibinfo {author} {\bibfnamefont {C.}~\bibnamefont {Yang}}, \bibinfo {author}
  {\bibfnamefont {A.}~\bibnamefont {Leenstra}}, \bibinfo {author}
  {\bibfnamefont {B.}~\bibnamefont {De~Ronde}}, \bibinfo {author}
  {\bibfnamefont {J.}~\bibnamefont {Dehollain}}, \bibinfo {author}
  {\bibfnamefont {J.}~\bibnamefont {Muhonen}}, \bibinfo {author} {\bibfnamefont
  {F.}~\bibnamefont {Hudson}}, \bibinfo {author} {\bibfnamefont {K.~M.}\
  \bibnamefont {Itoh}}, \bibinfo {author} {\bibfnamefont {A.}~\bibnamefont
  {Morello}},  \emph {et~al.},\ }\href@noop {} {\bibfield  {journal} {\bibinfo
  {journal} {Nature nanotechnology}\ }\textbf {\bibinfo {volume} {9}},\
  \bibinfo {pages} {981} (\bibinfo {year} {2014})}\BibitemShut {NoStop}%
\bibitem [{\citenamefont {Studenikin}\ \emph {et~al.}(2006)\citenamefont
  {Studenikin}, \citenamefont {Gaudreau}, \citenamefont {Sachrajda},
  \citenamefont {Zawadzki}, \citenamefont {Kam}, \citenamefont {Lapointe},
  \citenamefont {Korkusinski},\ and\ \citenamefont
  {Hawrylak}}]{Studenikin2006}%
  \BibitemOpen
  \bibfield  {author} {\bibinfo {author} {\bibfnamefont {S.}~\bibnamefont
  {Studenikin}}, \bibinfo {author} {\bibfnamefont {L.}~\bibnamefont
  {Gaudreau}}, \bibinfo {author} {\bibfnamefont {A.}~\bibnamefont {Sachrajda}},
  \bibinfo {author} {\bibfnamefont {P.}~\bibnamefont {Zawadzki}}, \bibinfo
  {author} {\bibfnamefont {A.}~\bibnamefont {Kam}}, \bibinfo {author}
  {\bibfnamefont {J.}~\bibnamefont {Lapointe}}, \bibinfo {author}
  {\bibfnamefont {M.}~\bibnamefont {Korkusinski}}, \ and\ \bibinfo {author}
  {\bibfnamefont {P.}~\bibnamefont {Hawrylak}},\ }in\ \href@noop {} {\emph
  {\bibinfo {booktitle} {Nanotechnology, 2006}}},\ Vol.~\bibinfo {volume} {2}\
  (\bibinfo {organization} {IEEE},\ \bibinfo {year} {2006})\ pp.\ \bibinfo
  {pages} {871--874}\BibitemShut {NoStop}%
\bibitem [{\citenamefont {Schr\"oer}\ \emph {et~al.}(2007)\citenamefont
  {Schr\"oer}, \citenamefont {Greentree}, \citenamefont {Gaudreau},
  \citenamefont {Eberl}, \citenamefont {Hollenberg}, \citenamefont {Kotthaus},\
  and\ \citenamefont {Ludwig}}]{Schroeer2007}%
  \BibitemOpen
  \bibfield  {author} {\bibinfo {author} {\bibfnamefont {D.}~\bibnamefont
  {Schr\"oer}}, \bibinfo {author} {\bibfnamefont {A.~D.}\ \bibnamefont
  {Greentree}}, \bibinfo {author} {\bibfnamefont {L.}~\bibnamefont {Gaudreau}},
  \bibinfo {author} {\bibfnamefont {K.}~\bibnamefont {Eberl}}, \bibinfo
  {author} {\bibfnamefont {L.~C.~L.}\ \bibnamefont {Hollenberg}}, \bibinfo
  {author} {\bibfnamefont {J.~P.}\ \bibnamefont {Kotthaus}}, \ and\ \bibinfo
  {author} {\bibfnamefont {S.}~\bibnamefont {Ludwig}},\ }\href {\doibase
  10.1103/PhysRevB.76.075306} {\bibfield  {journal} {\bibinfo  {journal} {Phys.
  Rev. B}\ }\textbf {\bibinfo {volume} {76}},\ \bibinfo {pages} {075306}
  (\bibinfo {year} {2007})}\BibitemShut {NoStop}%
\bibitem [{\citenamefont {Thalineau}\ \emph {et~al.}(2012)\citenamefont
  {Thalineau}, \citenamefont {Hermelin}, \citenamefont {Wieck}, \citenamefont
  {B\"auerle}, \citenamefont {Saminadayar},\ and\ \citenamefont
  {Meunier}}]{Thalineau2012}%
  \BibitemOpen
  \bibfield  {author} {\bibinfo {author} {\bibfnamefont {R.}~\bibnamefont
  {Thalineau}}, \bibinfo {author} {\bibfnamefont {S.}~\bibnamefont {Hermelin}},
  \bibinfo {author} {\bibfnamefont {A.~D.}\ \bibnamefont {Wieck}}, \bibinfo
  {author} {\bibfnamefont {C.}~\bibnamefont {B\"auerle}}, \bibinfo {author}
  {\bibfnamefont {L.}~\bibnamefont {Saminadayar}}, \ and\ \bibinfo {author}
  {\bibfnamefont {T.}~\bibnamefont {Meunier}},\ }\href {\doibase
  10.1063/1.4749811} {\bibfield  {journal} {\bibinfo  {journal} {Appl. Phys.
  Lett.}\ }\textbf {\bibinfo {volume} {101}},\ \bibinfo {eid} {103102}
  (\bibinfo {year} {2012})}\BibitemShut {NoStop}%
\bibitem [{\citenamefont {Takakura}\ \emph {et~al.}(2014)\citenamefont
  {Takakura}, \citenamefont {Noiri}, \citenamefont {Obata}, \citenamefont
  {Otsuka}, \citenamefont {Yoneda}, \citenamefont {Yoshida},\ and\
  \citenamefont {Tarucha}}]{Takakura2014}%
  \BibitemOpen
  \bibfield  {author} {\bibinfo {author} {\bibfnamefont {T.}~\bibnamefont
  {Takakura}}, \bibinfo {author} {\bibfnamefont {A.}~\bibnamefont {Noiri}},
  \bibinfo {author} {\bibfnamefont {T.}~\bibnamefont {Obata}}, \bibinfo
  {author} {\bibfnamefont {T.}~\bibnamefont {Otsuka}}, \bibinfo {author}
  {\bibfnamefont {J.}~\bibnamefont {Yoneda}}, \bibinfo {author} {\bibfnamefont
  {K.}~\bibnamefont {Yoshida}}, \ and\ \bibinfo {author} {\bibfnamefont
  {S.}~\bibnamefont {Tarucha}},\ }\href {\doibase 10.1063/1.4869108} {\bibfield
   {journal} {\bibinfo  {journal} {Appl. Phys. Lett.}\ }\textbf {\bibinfo
  {volume} {104}},\ \bibinfo {eid} {113109} (\bibinfo {year}
  {2014})}\BibitemShut {NoStop}%
\bibitem [{\citenamefont {Eng}\ \emph {et~al.}(2015)\citenamefont {Eng},
  \citenamefont {Ladd}, \citenamefont {Smith}, \citenamefont {Borselli},
  \citenamefont {Kiselev}, \citenamefont {Fong}, \citenamefont {Holabird},
  \citenamefont {Hazard}, \citenamefont {Huang}, \citenamefont {Deelman},
  \citenamefont {Milosavljevic}, \citenamefont {Schmitz}, \citenamefont {Ross},
  \citenamefont {Gyure},\ and\ \citenamefont {Hunter}}]{Eng2015}%
  \BibitemOpen
  \bibfield  {author} {\bibinfo {author} {\bibfnamefont {K.}~\bibnamefont
  {Eng}}, \bibinfo {author} {\bibfnamefont {T.~D.}\ \bibnamefont {Ladd}},
  \bibinfo {author} {\bibfnamefont {A.}~\bibnamefont {Smith}}, \bibinfo
  {author} {\bibfnamefont {M.~G.}\ \bibnamefont {Borselli}}, \bibinfo {author}
  {\bibfnamefont {A.~A.}\ \bibnamefont {Kiselev}}, \bibinfo {author}
  {\bibfnamefont {B.~H.}\ \bibnamefont {Fong}}, \bibinfo {author}
  {\bibfnamefont {K.~S.}\ \bibnamefont {Holabird}}, \bibinfo {author}
  {\bibfnamefont {T.~M.}\ \bibnamefont {Hazard}}, \bibinfo {author}
  {\bibfnamefont {B.}~\bibnamefont {Huang}}, \bibinfo {author} {\bibfnamefont
  {P.~W.}\ \bibnamefont {Deelman}}, \bibinfo {author} {\bibfnamefont
  {I.}~\bibnamefont {Milosavljevic}}, \bibinfo {author} {\bibfnamefont {A.~E.}\
  \bibnamefont {Schmitz}}, \bibinfo {author} {\bibfnamefont {R.~S.}\
  \bibnamefont {Ross}}, \bibinfo {author} {\bibfnamefont {M.~F.}\ \bibnamefont
  {Gyure}}, \ and\ \bibinfo {author} {\bibfnamefont {A.~T.}\ \bibnamefont
  {Hunter}},\ }\href@noop {} {\bibfield  {journal} {\bibinfo  {journal}
  {Science Advances}\ }\textbf {\bibinfo {volume} {1}} (\bibinfo {year}
  {2015})}\BibitemShut {NoStop}%
\bibitem [{\citenamefont {Baart}\ \emph {et~al.}(2016)\citenamefont {Baart},
  \citenamefont {Shafiei}, \citenamefont {Fujita}, \citenamefont {Reichl},
  \citenamefont {Wegscheider},\ and\ \citenamefont {L.~M.~K.}}]{Baart2016}%
  \BibitemOpen
  \bibfield  {author} {\bibinfo {author} {\bibfnamefont {T.~A.}\ \bibnamefont
  {Baart}}, \bibinfo {author} {\bibfnamefont {M.}~\bibnamefont {Shafiei}},
  \bibinfo {author} {\bibfnamefont {T.}~\bibnamefont {Fujita}}, \bibinfo
  {author} {\bibfnamefont {C.}~\bibnamefont {Reichl}}, \bibinfo {author}
  {\bibfnamefont {W.}~\bibnamefont {Wegscheider}}, \ and\ \bibinfo {author}
  {\bibfnamefont {V.}~\bibnamefont {L.~M.~K.}},\ }\href
  {http://dx.doi.org/10.1038/nnano.2015.291} {\bibfield  {journal} {\bibinfo
  {journal} {Nat. Nano}\ }\textbf {\bibinfo {volume} {11}},\ \bibinfo {pages}
  {330} (\bibinfo {year} {2016})}\BibitemShut {NoStop}%
\bibitem [{\citenamefont {Seo}\ \emph {et~al.}(2013)\citenamefont {Seo},
  \citenamefont {Choi}, \citenamefont {Lee}, \citenamefont {Kim}, \citenamefont
  {Chung}, \citenamefont {Sim}, \citenamefont {Umansky},\ and\ \citenamefont
  {Mahalu}}]{Seo2013}%
  \BibitemOpen
  \bibfield  {author} {\bibinfo {author} {\bibfnamefont {M.}~\bibnamefont
  {Seo}}, \bibinfo {author} {\bibfnamefont {H.~K.}\ \bibnamefont {Choi}},
  \bibinfo {author} {\bibfnamefont {S.-Y.}\ \bibnamefont {Lee}}, \bibinfo
  {author} {\bibfnamefont {N.}~\bibnamefont {Kim}}, \bibinfo {author}
  {\bibfnamefont {Y.}~\bibnamefont {Chung}}, \bibinfo {author} {\bibfnamefont
  {H.-S.}\ \bibnamefont {Sim}}, \bibinfo {author} {\bibfnamefont
  {V.}~\bibnamefont {Umansky}}, \ and\ \bibinfo {author} {\bibfnamefont
  {D.}~\bibnamefont {Mahalu}},\ }\href {\doibase
  10.1103/PhysRevLett.110.046803} {\bibfield  {journal} {\bibinfo  {journal}
  {Phys. Rev. Lett.}\ }\textbf {\bibinfo {volume} {110}},\ \bibinfo {pages}
  {046803} (\bibinfo {year} {2013})}\BibitemShut {NoStop}%
\bibitem [{\citenamefont {Laird}\ \emph {et~al.}(2010)\citenamefont {Laird},
  \citenamefont {Taylor}, \citenamefont {DiVincenzo}, \citenamefont {Marcus},
  \citenamefont {Hanson},\ and\ \citenamefont {Gossard}}]{Laird2010}%
  \BibitemOpen
  \bibfield  {author} {\bibinfo {author} {\bibfnamefont {E.~A.}\ \bibnamefont
  {Laird}}, \bibinfo {author} {\bibfnamefont {J.~M.}\ \bibnamefont {Taylor}},
  \bibinfo {author} {\bibfnamefont {D.~P.}\ \bibnamefont {DiVincenzo}},
  \bibinfo {author} {\bibfnamefont {C.~M.}\ \bibnamefont {Marcus}}, \bibinfo
  {author} {\bibfnamefont {M.~P.}\ \bibnamefont {Hanson}}, \ and\ \bibinfo
  {author} {\bibfnamefont {A.~C.}\ \bibnamefont {Gossard}},\ }\href {\doibase
  10.1103/PhysRevB.82.075403} {\bibfield  {journal} {\bibinfo  {journal} {Phys.
  Rev. B}\ }\textbf {\bibinfo {volume} {82}},\ \bibinfo {pages} {075403}
  (\bibinfo {year} {2010})}\BibitemShut {NoStop}%
\bibitem [{\citenamefont {Gaudreau}\ \emph {et~al.}(2012)\citenamefont
  {Gaudreau}, \citenamefont {Granger}, \citenamefont {Kam}, \citenamefont
  {Aers}, \citenamefont {Studenikin}, \citenamefont {Zawadzki}, \citenamefont
  {Pioro-Ladriere}, \citenamefont {Wasilewski},\ and\ \citenamefont
  {Sachrajda}}]{Gaudreau2012}%
  \BibitemOpen
  \bibfield  {author} {\bibinfo {author} {\bibfnamefont {L.}~\bibnamefont
  {Gaudreau}}, \bibinfo {author} {\bibfnamefont {G.}~\bibnamefont {Granger}},
  \bibinfo {author} {\bibfnamefont {A.}~\bibnamefont {Kam}}, \bibinfo {author}
  {\bibfnamefont {G.}~\bibnamefont {Aers}}, \bibinfo {author} {\bibfnamefont
  {S.}~\bibnamefont {Studenikin}}, \bibinfo {author} {\bibfnamefont
  {P.}~\bibnamefont {Zawadzki}}, \bibinfo {author} {\bibfnamefont
  {M.}~\bibnamefont {Pioro-Ladriere}}, \bibinfo {author} {\bibfnamefont
  {Z.}~\bibnamefont {Wasilewski}}, \ and\ \bibinfo {author} {\bibfnamefont
  {A.}~\bibnamefont {Sachrajda}},\ }\href@noop {} {\bibfield  {journal}
  {\bibinfo  {journal} {Nat. Phys.}\ }\textbf {\bibinfo {volume} {8}},\
  \bibinfo {pages} {54} (\bibinfo {year} {2012})}\BibitemShut {NoStop}%
\bibitem [{\citenamefont {Noiri}\ \emph {et~al.}(2016)\citenamefont {Noiri},
  \citenamefont {Yoneda}, \citenamefont {Nakajima}, \citenamefont {Otsuka},
  \citenamefont {Delbecq}, \citenamefont {Takeda}, \citenamefont {Amaha},
  \citenamefont {Allison}, \citenamefont {Ludwig}, \citenamefont {Wieck} \emph
  {et~al.}}]{noiri2016}%
  \BibitemOpen
  \bibfield  {author} {\bibinfo {author} {\bibfnamefont {A.}~\bibnamefont
  {Noiri}}, \bibinfo {author} {\bibfnamefont {J.}~\bibnamefont {Yoneda}},
  \bibinfo {author} {\bibfnamefont {T.}~\bibnamefont {Nakajima}}, \bibinfo
  {author} {\bibfnamefont {T.}~\bibnamefont {Otsuka}}, \bibinfo {author}
  {\bibfnamefont {M.~R.}\ \bibnamefont {Delbecq}}, \bibinfo {author}
  {\bibfnamefont {K.}~\bibnamefont {Takeda}}, \bibinfo {author} {\bibfnamefont
  {S.}~\bibnamefont {Amaha}}, \bibinfo {author} {\bibfnamefont
  {G.}~\bibnamefont {Allison}}, \bibinfo {author} {\bibfnamefont
  {A.}~\bibnamefont {Ludwig}}, \bibinfo {author} {\bibfnamefont {A.~D.}\
  \bibnamefont {Wieck}},  \emph {et~al.},\ }\href@noop {} {\bibfield  {journal}
  {\bibinfo  {journal} {Applied Physics Letters}\ }\textbf {\bibinfo {volume}
  {108}},\ \bibinfo {pages} {153101} (\bibinfo {year} {2016})}\BibitemShut
  {NoStop}%
\bibitem [{\citenamefont {Bertrand}\ \emph {et~al.}(2015)\citenamefont
  {Bertrand}, \citenamefont {Flentje}, \citenamefont {Takada}, \citenamefont
  {Yamamoto}, \citenamefont {Tarucha}, \citenamefont {Ludwig}, \citenamefont
  {Wieck}, \citenamefont {B\"auerle},\ and\ \citenamefont
  {Meunier}}]{Bertrand2015}%
  \BibitemOpen
  \bibfield  {author} {\bibinfo {author} {\bibfnamefont {B.}~\bibnamefont
  {Bertrand}}, \bibinfo {author} {\bibfnamefont {H.}~\bibnamefont {Flentje}},
  \bibinfo {author} {\bibfnamefont {S.}~\bibnamefont {Takada}}, \bibinfo
  {author} {\bibfnamefont {M.}~\bibnamefont {Yamamoto}}, \bibinfo {author}
  {\bibfnamefont {S.}~\bibnamefont {Tarucha}}, \bibinfo {author} {\bibfnamefont
  {A.}~\bibnamefont {Ludwig}}, \bibinfo {author} {\bibfnamefont {A.~D.}\
  \bibnamefont {Wieck}}, \bibinfo {author} {\bibfnamefont {C.}~\bibnamefont
  {B\"auerle}}, \ and\ \bibinfo {author} {\bibfnamefont {T.}~\bibnamefont
  {Meunier}},\ }\href {\doibase 10.1103/PhysRevLett.115.096801} {\bibfield
  {journal} {\bibinfo  {journal} {Phys. Rev. Lett.}\ }\textbf {\bibinfo
  {volume} {115}},\ \bibinfo {pages} {096801} (\bibinfo {year}
  {2015})}\BibitemShut {NoStop}%
\bibitem [{\citenamefont {Flentje}\ \emph {et~al.}(2017)\citenamefont
  {Flentje}, \citenamefont {Mortemousque}, \citenamefont {Thalineau},
  \citenamefont {Ludwig}, \citenamefont {Wieck}, \citenamefont {B{\"a}uerle},\
  and\ \citenamefont {Meunier}}]{flentje2017coherent}%
  \BibitemOpen
  \bibfield  {author} {\bibinfo {author} {\bibfnamefont {H.}~\bibnamefont
  {Flentje}}, \bibinfo {author} {\bibfnamefont {P.}~\bibnamefont
  {Mortemousque}}, \bibinfo {author} {\bibfnamefont {R.}~\bibnamefont
  {Thalineau}}, \bibinfo {author} {\bibfnamefont {A.}~\bibnamefont {Ludwig}},
  \bibinfo {author} {\bibfnamefont {A.}~\bibnamefont {Wieck}}, \bibinfo
  {author} {\bibfnamefont {C.}~\bibnamefont {B{\"a}uerle}}, \ and\ \bibinfo
  {author} {\bibfnamefont {T.}~\bibnamefont {Meunier}},\ }\href@noop {}
  {\bibfield  {journal} {\bibinfo  {journal} {arXiv preprint arXiv:1701.01279}\
  } (\bibinfo {year} {2017})}\BibitemShut {NoStop}%
\bibitem [{\citenamefont {van~der Wiel}\ \emph {et~al.}(2002)\citenamefont
  {van~der Wiel}, \citenamefont {De~Franceschi}, \citenamefont {Elzerman},
  \citenamefont {Fujisawa}, \citenamefont {Tarucha},\ and\ \citenamefont
  {Kouwenhoven}}]{RevModPhys.75.1}%
  \BibitemOpen
  \bibfield  {author} {\bibinfo {author} {\bibfnamefont {W.~G.}\ \bibnamefont
  {van~der Wiel}}, \bibinfo {author} {\bibfnamefont {S.}~\bibnamefont
  {De~Franceschi}}, \bibinfo {author} {\bibfnamefont {J.~M.}\ \bibnamefont
  {Elzerman}}, \bibinfo {author} {\bibfnamefont {T.}~\bibnamefont {Fujisawa}},
  \bibinfo {author} {\bibfnamefont {S.}~\bibnamefont {Tarucha}}, \ and\
  \bibinfo {author} {\bibfnamefont {L.~P.}\ \bibnamefont {Kouwenhoven}},\
  }\href {\doibase 10.1103/RevModPhys.75.1} {\bibfield  {journal} {\bibinfo
  {journal} {Rev. Mod. Phys.}\ }\textbf {\bibinfo {volume} {75}},\ \bibinfo
  {pages} {1} (\bibinfo {year} {2002})}\BibitemShut {NoStop}%
\bibitem [{\citenamefont {Davies}, \citenamefont {Larkin},\ and\ \citenamefont
  {Sukhorukov}(1995)}]{Davies1995}%
  \BibitemOpen
  \bibfield  {author} {\bibinfo {author} {\bibfnamefont {J.~H.}\ \bibnamefont
  {Davies}}, \bibinfo {author} {\bibfnamefont {I.~A.}\ \bibnamefont {Larkin}},
  \ and\ \bibinfo {author} {\bibfnamefont {E.~V.}\ \bibnamefont {Sukhorukov}},\
  }\href {\doibase 10.1063/1.359446} {\bibfield  {journal} {\bibinfo  {journal}
  {Journal of Applied Physics}\ }\textbf {\bibinfo {volume} {77}},\ \bibinfo
  {pages} {4504} (\bibinfo {year} {1995})}\BibitemShut {NoStop}%
\bibitem [{\citenamefont {Bautze}\ \emph {et~al.}(2014)\citenamefont {Bautze},
  \citenamefont {S\"ussmeier}, \citenamefont {Takada}, \citenamefont {Groth},
  \citenamefont {Meunier}, \citenamefont {Yamamoto}, \citenamefont {Tarucha},
  \citenamefont {Waintal},\ and\ \citenamefont {B\"auerle}}]{Bautze2014}%
  \BibitemOpen
  \bibfield  {author} {\bibinfo {author} {\bibfnamefont {T.}~\bibnamefont
  {Bautze}}, \bibinfo {author} {\bibfnamefont {C.}~\bibnamefont {S\"ussmeier}},
  \bibinfo {author} {\bibfnamefont {S.}~\bibnamefont {Takada}}, \bibinfo
  {author} {\bibfnamefont {C.}~\bibnamefont {Groth}}, \bibinfo {author}
  {\bibfnamefont {T.}~\bibnamefont {Meunier}}, \bibinfo {author} {\bibfnamefont
  {M.}~\bibnamefont {Yamamoto}}, \bibinfo {author} {\bibfnamefont
  {S.}~\bibnamefont {Tarucha}}, \bibinfo {author} {\bibfnamefont
  {X.}~\bibnamefont {Waintal}}, \ and\ \bibinfo {author} {\bibfnamefont
  {C.}~\bibnamefont {B\"auerle}},\ }\href {\doibase 10.1103/PhysRevB.89.125432}
  {\bibfield  {journal} {\bibinfo  {journal} {Phys. Rev. B}\ }\textbf {\bibinfo
  {volume} {89}},\ \bibinfo {pages} {125432} (\bibinfo {year}
  {2014})}\BibitemShut {NoStop}%
\bibitem [{\citenamefont {Field}\ \emph {et~al.}(1993)\citenamefont {Field},
  \citenamefont {Smith}, \citenamefont {Pepper}, \citenamefont {Ritchie},
  \citenamefont {Frost}, \citenamefont {Jones},\ and\ \citenamefont
  {Hasko}}]{Field1993}%
  \BibitemOpen
  \bibfield  {author} {\bibinfo {author} {\bibfnamefont {M.}~\bibnamefont
  {Field}}, \bibinfo {author} {\bibfnamefont {C.~G.}\ \bibnamefont {Smith}},
  \bibinfo {author} {\bibfnamefont {M.}~\bibnamefont {Pepper}}, \bibinfo
  {author} {\bibfnamefont {D.~A.}\ \bibnamefont {Ritchie}}, \bibinfo {author}
  {\bibfnamefont {J.~E.~F.}\ \bibnamefont {Frost}}, \bibinfo {author}
  {\bibfnamefont {G.~A.~C.}\ \bibnamefont {Jones}}, \ and\ \bibinfo {author}
  {\bibfnamefont {D.~G.}\ \bibnamefont {Hasko}},\ }\href {\doibase
  10.1103/PhysRevLett.70.1311} {\bibfield  {journal} {\bibinfo  {journal}
  {Phys. Rev. Lett.}\ }\textbf {\bibinfo {volume} {70}},\ \bibinfo {pages}
  {1311} (\bibinfo {year} {1993})}\BibitemShut {NoStop}%
\bibitem [{\citenamefont {Foxman}\ \emph {et~al.}(1994)\citenamefont {Foxman},
  \citenamefont {Meirav}, \citenamefont {McEuen}, \citenamefont {Kastner},
  \citenamefont {Klein}, \citenamefont {Belk}, \citenamefont {Abusch},\ and\
  \citenamefont {Wind}}]{Foxman1994}%
  \BibitemOpen
  \bibfield  {author} {\bibinfo {author} {\bibfnamefont {E.~B.}\ \bibnamefont
  {Foxman}}, \bibinfo {author} {\bibfnamefont {U.}~\bibnamefont {Meirav}},
  \bibinfo {author} {\bibfnamefont {P.~L.}\ \bibnamefont {McEuen}}, \bibinfo
  {author} {\bibfnamefont {M.~A.}\ \bibnamefont {Kastner}}, \bibinfo {author}
  {\bibfnamefont {O.}~\bibnamefont {Klein}}, \bibinfo {author} {\bibfnamefont
  {P.~A.}\ \bibnamefont {Belk}}, \bibinfo {author} {\bibfnamefont {D.~M.}\
  \bibnamefont {Abusch}}, \ and\ \bibinfo {author} {\bibfnamefont {S.~J.}\
  \bibnamefont {Wind}},\ }\href {\doibase 10.1103/PhysRevB.50.14193} {\bibfield
   {journal} {\bibinfo  {journal} {Phys. Rev. B}\ }\textbf {\bibinfo {volume}
  {50}},\ \bibinfo {pages} {14193} (\bibinfo {year} {1994})}\BibitemShut
  {NoStop}%
\bibitem [{\citenamefont {Meir}, \citenamefont {Wingreen},\ and\ \citenamefont
  {Lee}(1991)}]{Meir1991}%
  \BibitemOpen
  \bibfield  {author} {\bibinfo {author} {\bibfnamefont {Y.}~\bibnamefont
  {Meir}}, \bibinfo {author} {\bibfnamefont {N.~S.}\ \bibnamefont {Wingreen}},
  \ and\ \bibinfo {author} {\bibfnamefont {P.~A.}\ \bibnamefont {Lee}},\ }\href
  {\doibase 10.1103/PhysRevLett.66.3048} {\bibfield  {journal} {\bibinfo
  {journal} {Phys. Rev. Lett.}\ }\textbf {\bibinfo {volume} {66}},\ \bibinfo
  {pages} {3048} (\bibinfo {year} {1991})}\BibitemShut {NoStop}%
\bibitem [{\citenamefont {DiCarlo}\ \emph {et~al.}(2004)\citenamefont
  {DiCarlo}, \citenamefont {Lynch}, \citenamefont {Johnson}, \citenamefont
  {Childress}, \citenamefont {Crockett}, \citenamefont {Marcus}, \citenamefont
  {Hanson},\ and\ \citenamefont {Gossard}}]{DiCarlo2004}%
  \BibitemOpen
  \bibfield  {author} {\bibinfo {author} {\bibfnamefont {L.}~\bibnamefont
  {DiCarlo}}, \bibinfo {author} {\bibfnamefont {H.~J.}\ \bibnamefont {Lynch}},
  \bibinfo {author} {\bibfnamefont {A.~C.}\ \bibnamefont {Johnson}}, \bibinfo
  {author} {\bibfnamefont {L.~I.}\ \bibnamefont {Childress}}, \bibinfo {author}
  {\bibfnamefont {K.}~\bibnamefont {Crockett}}, \bibinfo {author}
  {\bibfnamefont {C.~M.}\ \bibnamefont {Marcus}}, \bibinfo {author}
  {\bibfnamefont {M.~P.}\ \bibnamefont {Hanson}}, \ and\ \bibinfo {author}
  {\bibfnamefont {A.~C.}\ \bibnamefont {Gossard}},\ }\href {\doibase
  10.1103/PhysRevLett.92.226801} {\bibfield  {journal} {\bibinfo  {journal}
  {Phys. Rev. Lett.}\ }\textbf {\bibinfo {volume} {92}},\ \bibinfo {pages}
  {226801} (\bibinfo {year} {2004})}\BibitemShut {NoStop}%
\bibitem [{\citenamefont {Braakman}\ \emph {et~al.}(2013)\citenamefont
  {Braakman}, \citenamefont {Barthelemy}, \citenamefont {Reichl}, \citenamefont
  {Wegscheider},\ and\ \citenamefont {Vandersypen}}]{braakman2013long}%
  \BibitemOpen
  \bibfield  {author} {\bibinfo {author} {\bibfnamefont {F.~R.}\ \bibnamefont
  {Braakman}}, \bibinfo {author} {\bibfnamefont {P.}~\bibnamefont
  {Barthelemy}}, \bibinfo {author} {\bibfnamefont {C.}~\bibnamefont {Reichl}},
  \bibinfo {author} {\bibfnamefont {W.}~\bibnamefont {Wegscheider}}, \ and\
  \bibinfo {author} {\bibfnamefont {L.~M.}\ \bibnamefont {Vandersypen}},\
  }\href@noop {} {\bibfield  {journal} {\bibinfo  {journal} {Nature
  nanotechnology}\ }\textbf {\bibinfo {volume} {8}},\ \bibinfo {pages} {432}
  (\bibinfo {year} {2013})}\BibitemShut {NoStop}%
\bibitem [{\citenamefont {Van~Houten}, \citenamefont {Beenakker},\ and\
  \citenamefont {Staring}(1992)}]{VanHouten1992}%
  \BibitemOpen
  \bibfield  {author} {\bibinfo {author} {\bibfnamefont {H.}~\bibnamefont
  {Van~Houten}}, \bibinfo {author} {\bibfnamefont {C.}~\bibnamefont
  {Beenakker}}, \ and\ \bibinfo {author} {\bibfnamefont {A.}~\bibnamefont
  {Staring}},\ }in\ \href@noop {} {\emph {\bibinfo {booktitle} {Single charge
  tunneling}}}\ (\bibinfo  {publisher} {Springer},\ \bibinfo {year} {1992})\
  pp.\ \bibinfo {pages} {167--216}\BibitemShut {NoStop}%
\bibitem [{\citenamefont {Oosterkamp}\ \emph {et~al.}(1998)\citenamefont
  {Oosterkamp}, \citenamefont {Fujisawa}, \citenamefont {Van Der~Wiel},
  \citenamefont {Ishibashi}, \citenamefont {Hijman}, \citenamefont {Tarucha},\
  and\ \citenamefont {Kouwenhoven}}]{Oosterkamp1998}%
  \BibitemOpen
  \bibfield  {author} {\bibinfo {author} {\bibfnamefont {T.}~\bibnamefont
  {Oosterkamp}}, \bibinfo {author} {\bibfnamefont {T.}~\bibnamefont
  {Fujisawa}}, \bibinfo {author} {\bibfnamefont {W.}~\bibnamefont {Van
  Der~Wiel}}, \bibinfo {author} {\bibfnamefont {K.}~\bibnamefont {Ishibashi}},
  \bibinfo {author} {\bibfnamefont {R.}~\bibnamefont {Hijman}}, \bibinfo
  {author} {\bibfnamefont {S.}~\bibnamefont {Tarucha}}, \ and\ \bibinfo
  {author} {\bibfnamefont {L.~P.}\ \bibnamefont {Kouwenhoven}},\ }\href@noop {}
  {\bibfield  {journal} {\bibinfo  {journal} {Nature}\ }\textbf {\bibinfo
  {volume} {395}},\ \bibinfo {pages} {873} (\bibinfo {year}
  {1998})}\BibitemShut {NoStop}%
\end{thebibliography}
\end{document}